\documentclass[a4paper]{article}
\usepackage[a4paper,top=80pt,lmargin=80pt,rmargin=80pt,bottom=80pt]{geometry}

\usepackage{graphicx}

\usepackage[main=english,croatian]{babel}

\usepackage[]{cite}

\expandafter\let\csname equation*\endcsname\relax
\expandafter\let\csname endequation*\endcsname\relax

\usepackage{tikz}
\newcommand{\btkz}{\begin{tikzpicture}}
\newcommand{\etkz}{\end{tikzpicture}}
\tikzset{
    partial ellipse/.style args={#1:#2:#3}{
        insert path={+ (#1:#3) arc (#1:#2:#3)}
    }
}

\usepackage[symbol]{footmisc}
\usepackage{amsmath}
\usepackage[]{amssymb}
\usepackage[]{mathrsfs}
\usepackage[]{eucal}
\usepackage[]{tensor}
\usepackage[]{amsthm}
\usepackage[]{bm}

\newcommand{\B}{\bm{B}}
\newcommand{\M}{\bm{M}}
\newcommand{\HH}{\bm{H}}
\newcommand{\un}{\bm{n}}
\newcommand{\vr}{\bm{r}}

\newcommand{\dd}{\partial}
\newcommand{\df}{\mathrm{d}}

\newcommand{\veps}{\varepsilon}

\newcommand{\qqd}{\ , \quad}
\newcommand{\bc}{\begin{center}}
\newcommand{\ec}{\end{center}}
\newcommand{\be}{\begin{equation}}
\newcommand{\ee}{\end{equation}}

\newcommand{\lb}{\left<}
\newcommand{\rb}{\right>}

\newcommand{\fun}[1]{\mathrm{#1}}

\newcommand{\norm}[1]{\left\lVert #1 \right\rVert}

\newcommand{\defeq}{\mathrel{\mathop:}=}

\usepackage{dsfont}

\newcommand{\rr}{\mathds{R}}

\usepackage[unicode]{hyperref}
\usepackage{xcolor}
\definecolor{pastgreen}{HTML}{669900}
\definecolor{pastblue}{HTML}{336699}
\definecolor{pastred}{HTML}{990000}
\definecolor{linkcol}{HTML}{663333}
\hypersetup{colorlinks,linkcolor={pastblue},citecolor={pastblue},urlcolor={pastblue}}

\theoremstyle{plain} \newtheorem{tm}{Theorem}[section]
\theoremstyle{plain} \newtheorem{lm}[tm]{Lemma}
\theoremstyle{plain} \newtheorem{defn}[tm]{Definition}
\newcommand{\btm}{\begin{tm}}
\newcommand{\etm}{\end{tm}}
\newcommand{\blm}{\begin{lm}}
\newcommand{\elm}{\end{lm}}
\newcommand{\bdefn}{\begin{defn}}
\newcommand{\edefn}{\end{defn}}

\begin{document}

\begin{flushright}
\texttt{ZTF-EP-24-09}
\end{flushright}

\vspace{20pt}

\bc
{\LARGE \textbf{Generalized Runcorn's theorem}}

\medskip

{\LARGE \textbf{and crustal magnetism}}

\vspace{20pt}

{\large Ivica Smoli\'c\footnote[1]{E-mail address: ismolic@phy.hr}}

\vspace{10pt}

Department of Physics, Faculty of Science, University of Zagreb, 

Bijeni\v cka cesta 32, 10000 Zagreb, Croatia
\ec

\vspace{20pt}

\textbf{Abstract.} During the era of NASA's Apollo missions, Keith S.~Runcorn proposed an explanation of discrepancy between the Moon's negligible global magnetic field and magnetized samples of lunar regolith, based on identical vanishing of external magnetic field of a spherical shell, magnetized by an internal source which is no longer present. We revisit and generalize the Runcorn's result, showing that it is a consequence of a (weighted) orthogonality of gradients of harmonic functions on a spherical shell in arbitrary number of dimensions. Furthermore, we explore bounds on external magnetic field in the case when the idealized spherical shell is replaced with a more realistic geometric shape and when the thermoremanent magnetization susceptibility deviates from the spherical symmetry. Finally, we analyse a model of thermoremanent magnetization acquired by crustal inward cooling of a spherical astrophysical body and put some general bounds on the associated magnetic field.

\vspace{20pt}

%%%%%%%%%%%%%%%%%%%%%%%%%%
\section{Introduction} %%%
%%%%%%%%%%%%%%%%%%%%%%%%%%

Magnetic field permeates all matter in Solar System, offering important insight into the history and internal structure of astrophysical bodies. Above all, here we encounter solar and planetary dynamos, movement of electrically conducting fluid described by magnetohydrodynamics. Also, even if a terrestrial planet or a moon no longer has an active dynamo in its core (or, indeed, never had one), it still may possess an imprint of the past magnetic fields in a form of the remanent magnetization in its outer, solid layers.

\smallskip

A particularly intriguing and instructive episode revolves around Earth's natural satellite, the Moon. Early investigations of lunar magnetism were conducted with orbital probes \emph{Luna 2} \cite{DYZP61} in 1959, \emph{Luna 10} \cite{DYZP66} in 1966 and \emph{Explorer 35} \cite{SCC67} in 1967, none of which detected global lunar magnetic field above the sensitivity of their magnetometers, with the order of magnitude ranging between 10 nT and 0.1 nT. These observations were later corroborated by orbital measurements from the \emph{Apollo 15} mission in 1971 and the \emph{Apollo 16} mission in 1972 (cf.~\cite{Fuller74,Russell80,LSB} and references therein), putting an upper bound on lunar magnetic dipole moment 8 orders of magnitude below the present Earth's magnetic dipole moment. On the other hand, analysis of the samples returned by the \emph{Apollo} missions have revealed that lunar rocks possess natural remanent magnetization, carried by iron particles, providing us with abundant evidence for the past lunar magnetic field.

\smallskip

Keith S.~Runcorn \cite{Runcorn75a,Runcorn75b,Runcorn77} has proposed an explanation of the seemingly inconsistent lunar magnetism\footnote[2]{An overview of theories for the origin of lunar magnetism proposed during 1970s is given in \cite{DD79}.}, based on the following simple, lucid observation. If a spherical shell bears a permanent magnetization proportional to a magnetic field, sourced by currents contained within the inner sphere, then the resulting magnetic field out of the outer sphere, in the absence of the source currents, is exactly zero. Based on data from several seismometers, placed on lunar surface during Apollo missions \cite{Wieczorek09}, we know that the Moon is a differentiated body, with crust, mantle and core. Runcorn's scenario suggests that the Moon once had active dynamo in its core, whose magnetic field has magnetized lunar crust; as the lunar dynamo is no longer active, external magnetic field of the remanent crustal magnetization is zero. Unfortunately, as was discussed already by Runcorn himself \cite{Runcorn75b,Runcorn77}, this model offers only a partial solution of the puzzle. First of all, the Moon is not ideally spherical, albeit one does not expect dramatic change of the conclusion as the Moon's deviation from spherical symmetry is of the order of $10^{-3}$ \cite{Runcorn77}. Secondly, although Runcorn allows for a radially dependent proportionality between magnetization and core's magnetic field, it does not take into account that outer layers will additionally magnetize inner layers (consequence of which was investigated by Srnka \cite{Srnka76,Srnka76p}). Overall, these amendments result in a limited global external magnetic field. However, further intricacies of the problem proliferate with localised magnetic fields, usually referred to as ``magnetic anomalies'', reaching the order of hundreds of nT. These were originally observed in \emph{Apollo} missions, through orbital and \emph{in situ} measurements, and more recently mapped in detail by \emph{Lunar Prospector} mission \cite{LP08} (for a discussion about the possible relation between lunar magnetic anomalies and impact processes cf.~\cite{WWS12,OWSMT20,Wetal23,YW24}). Finally, although the consensus in the community seems to be that the Moon once had long-lived dynamo \cite{WT14,OWSMT20,TE22,Wetal23}, this conclusion was put under question with more recent reassessment of the lunar samples \cite{Tarduno21}. Nevertheless, it is essential to stress that even if the Runcorn's basic argument is insufficient to explain \emph{lunar} magnetic field, it still remains a valuable idea which might be applied to some other astrophysical bodies, as was previously shown in the case of the Hermean\footnote{Modern consensus, based on all available data (cf.~\cite{MercuryBook} and references therein), is that a core dynamo is favoured over remanent crustal magnetization as the main candidate source of the Mercury's magnetic field.} \cite{Stephenson76b} and Martian \cite{LS97} magnetic fields. 

\smallskip

Looking back at the original Runcorn's theorem, we are led to a number of glaring questions. Where does this result ``come from'', namely can it be reformulated as a corollary to some more general property of an appropriately chosen family of functions? What happens with the conclusions once the geometry of the problem is ``slightly perturbed''? What can be deduced from a model which takes into account magnetization between the layers of the crust? We shall examine these questions more closely and provide some insights which could be useful for future investigations of the crustal magnetism. The paper is organized as follows. In section 2 we state and prove a generalized Runcorn's theorem, a property of harmonic functions. In section 3 we analyse magnetic field of a magnetized crust in the cases when some of the Runcorn's assumptions are not met. In section 4 we revisit Srnka's model of a crust, magnetized during the inward cooling, and look more closely into the pertaining bounds on the magnetic field.

\medskip

\emph{Notation and conventions}. Throughout the paper we use SI units. Euclidean norm is denoted by $\norm{\,.\,}$ (without additional indices), while $L^p$ norm is denoted by $\norm{\,.\,}_{L^p}$. Sums which appear in decomposition of harmonic functions with spherical harmonics are abbreviated as $\sum_{\ell,m} \defeq \sum_{\ell=0}^\infty \sum_{m=-\ell}^\ell$.

%%%%%%%%%%%%%%%%%%%%%%%%%%%%%%%%%%%%%%%%%%%
\section{Generalized Runcorn's theorem} %%%
%%%%%%%%%%%%%%%%%%%%%%%%%%%%%%%%%%%%%%%%%%%

Context of the Runcorn's original theorem is magnetostatics, so let us first briefly recapitulate main equations. At each point of the space with known magnetic field $\B$ and magnetization $\M$, we can introduce an auxiliary vector field
\be
\HH \defeq \frac{1}{\mu_0} \, \B - \M ,
\ee
whose curl is equal to the density of the free electric current, $\nabla\times\HH = \bm{J}_\mathrm{f}$. On a simply connected domain in which the free currents are absent, we can introduce the magnetic scalar potential $\Psi$ via
\be
\HH = -\nabla\Psi ,
\ee
which is a solution of the Poisson's equation,
\be
\Delta\Psi = \nabla\cdot\M .
\ee
Now, suppose that free currents are completely absent in the space, while magnetized matter fills a domain $\Omega \subseteq \rr^3$, nonempty bounded open set with smooth boundary $\dd\Omega$. Then the magnetic scalar potential may be written (cf.~chapter 5.9 in \cite{Jack}), at least for all $\vr \in \rr^3 - \overline{\Omega}$, as
\be\label{eq:altPsi}
\Psi(\vr) = \frac{1}{4\pi} \left( - \int_\Omega \frac{\nabla'\cdot\M(\vr')}{\norm{\vr-\vr'}} \, \df V' + \oint_{\dd\Omega} \frac{\M(\vr')\cdot\un}{\norm{\vr-\vr'}} \, \df a' \right)
\ee
with the outward pointing normal $\bm{n}$ to $\dd\Omega$ or, using the divergence theorem, as
\be\label{eq:intPsi}
\Psi(\vr) = \frac{1}{4\pi} \int_\Omega \M(\vr') \cdot \nabla' \frac{1}{\norm{\vr-\vr'}} \, \df V' .
\ee
More concretely, let $\Omega$ be a spherical shell (``crust'' of the astrophysical body), bounded by concentric spheres of radii $0 < a < b$, while matter was magnetized with field $\HH_0$, sourced by currents contained in ball $r < a$ (which are no longer present). As the exterior $r > b$ is simply connected and, by assumption, devoid of free currents, we may write $\HH_0 = -\nabla \Psi_0$, with scalar potential $\Psi_0$ which is a solution of the Laplace's equation, $\Delta\Psi_0 = 0$, and vanishes at infinity. Crucial physical assumption is that magnetization is linear in the sense that $\M = \kappa \HH_0$ with some function $\kappa$, that is,
\be
\M(\vr) = -\kappa(\vr)\nabla\Psi_0(\vr) .
\ee
Depending on the context, function $\kappa$ may be ordinary magnetic susceptibility $\chi_m$, the thermoremanent magnetization susceptibility $\chi_\mathrm{TRM}$ \cite{WT14,TE22} or some combination of both, but we may, for simplicity, refer to it simply as a ``susceptibility''.

\smallskip

Runcorn has proven\footnote[3]{Proofs for constant $\kappa$ in papers \cite{Runcorn75a} and \cite{Runcorn77} are based on convenient treatment of the integral (\ref{eq:intPsi}), while the proof in \cite{Runcorn75b} relies on solving of the junction conditions. Generalization for radially dependent susceptibility $\kappa$ is presented in \cite{Runcorn75b} for dipole magnetizing field $\HH_0$.} that for isotropic $\kappa = \kappa(r)$ we have $\Psi(\vr) = 0$ for all $\norm{\vr} > b$. Our first goal is to generalize this result and place it into a wider context. The fact that both functions $\Psi_0$ and $\norm{\vr-\vr'}^{-1}$ are solutions of the Laplace's equation on the intersection of their domains, clearly indicates that one must look at the properties of the harmonic functions \cite{ABR}. In order to demonstrate that the original Runcorn's theorem is not just some peculiarity that occurs only in 3-dimensional spaces, we shall present the generalization to an arbitrary number of dimensions.

\btm\label{tm:GRtm}
Let $\mathscr{U} = \{ \bm{x}\in\rr^m \mid \norm{\bm{x}} < r_+ \}$ and $\mathscr{V} = \{ \bm{x}\in\rr^m \mid \norm{\bm{x}} > r_- \}$ be open sets, with $m \ge 2$ and radii $0 < r_- < r_+$. Furthermore, let $u$ be a harmonic function on $\mathscr{U}$, $v$ a harmonic function on $\mathscr{V}$, vanishing at infinity, $\lim_{\norm{\bm{x}} \to\infty} v(\bm{x}) = 0$, and $f : \lb 0,\infty \rb \to \rr$ a $C^1$ function. Then for any $r_- < a < b < r_+$ and a spherical shell $\mathcal{A} = \{ \bm{x} \in \rr^m \mid a \le \norm{\bm{x}} \le b \}$, the following orthogonality holds:
\be\label{eq:GR}
\int_{\mathcal{A}} f(\norm{\bm{x}}) \nabla u(\bm{x}) \cdot \nabla v(\bm{x}) \, \df V = 0 .
\ee
\etm

\smallskip

\noindent
\emph{Proof}. We shall introduce the radial coordinate $r \defeq \norm{\bm{x}}$ and orient the  boundary $\dd\mathcal{A}$ with the outward pointing unit vector field $\un$ (i.e.~$\un = \bm{\hat{r}}$ on $r = b$ and $\un = -\bm{\hat{r}}$ on $r = a$). Integration by parts and the generalized Stokes' theorem lead us to
\be
\int_{\mathcal{A}} f(r) \nabla u(\bm{x}) \cdot \nabla v(\bm{x}) \, \df V = \oint_{\dd\mathcal{A}} \!f(r) v(\bm{x}) D_n u(\bm{x}) \, \df a \, - \int_{\mathcal{A}} f'(r) v(\bm{x}) \dd_r u(\bm{x}) \, \df V ,
\ee
with the abbreviation $D_n \defeq \un\cdot\nabla$. Now, the backbone of the proof is the expansion of harmonic functions in terms of homogeneous harmonic polynomials, homogeneous polynomials which are solutions of the Laplace's equation. Results from chapter 10 in \cite{ABR} (theorem 10.1 for $m \ge 3$ and exercise 10.1 for $m = 2$) imply that harmonic functions $u$ and $v$, satisfying assumptions from the theorem, may be decomposed, each on its corresponding domain, as
\be
u(\bm{x}) = \sum_{k=0}^\infty p_k(\bm{x}) \qqd v(\bm{x}) = \sum_{\ell=0}^\infty \frac{q_\ell(\bm{x})}{\norm{\bm{x}}^{2\ell + m-2}} \, ,
\ee
where $p_k$ and $q_\ell$ are harmonic polynomials of degree, respectfully, $k$ and $\ell$. These series converge absolutely and uniformly on any compact subset of $\mathcal{A}$, allowing us to perform term-by-term derivations and the exchange of integrals with the sums.

\smallskip

In preparation for the evaluation of the integrals, we stress several elementary observations. Each $\bm{x} \ne \bm{0}$ can be decomposed as $\bm{x} = r\bm{s}$, with $\bm{s} \in S \defeq \{\bm{y} \in \rr^m \mid \norm{\bm{y}} = 1 \}$. Homogeneity of the harmonic polynomials implies that $p_k(\bm{x}) = p_k(r\bm{s}) = r^k p_k(\bm{s})$ and $\dd_r p_k(\bm{x}) = \dd_r (r^k p_k(\bm{s})) = kr^{k-1} p_k(\bm{s})$. Furthermore, if we denote by $\df\sigma$ surface element on the unit sphere $S$, then the surface and the volume elements may be decomposed, respectfully, as $\df a = r^{m-1} \, \df\sigma$ and $\df V = r^{m-1} \df r \, \df \sigma$. Finally, harmonic polynomials are orthogonal on the unit sphere $S$ in the sense that (see e.g.~proposition 5.9 in \cite{ABR})
\be
\oint_S q_\ell(\bm{s}) p_k(\bm{s}) \, \df \sigma = 0
\ee
for $k \ne \ell$. Hence,
\begin{align*}
\oint_{\dd\mathcal{A}} f(r) \, \frac{q_\ell(\bm{x})}{r^{2\ell + m-2}} \, D_n p_k(\bm{x}) \, \df a & = k(f(b)b^{k-\ell} - f(a)a^{k-\ell}) \oint_S q_\ell(\bm{s}) p_k(\bm{s}) \, \df \sigma \\
 & = k(f(b) - f(a)) \delta_{k\ell} \oint_S q_\ell(\bm{s}) p_k(\bm{s}) \, \df \sigma ,
\end{align*}
and
\begin{align*}
\int_\mathcal{A} f'(r) \, \frac{q_\ell(\bm{x})}{r^{2\ell + m-2}} \, \dd_r p_k(\bm{x}) \, \df V & = k \int_a^b f'(r) r^{k-\ell} \, \df r \oint_S q_\ell(\bm{s}) p_k(\bm{s}) \, \df \sigma \\
 & = k \delta_{k\ell} \oint_S q_\ell(\bm{s}) p_k(\bm{s}) \, \df \sigma \int_a^b f'(r) \, \df r \\
 & = k(f(b) - f(a)) \delta_{k\ell} \oint_S q_\ell(\bm{s}) p_k(\bm{s}) \, \df \sigma .
\end{align*}
The claim immediately follows by subtraction of the two integrals. We note in passing that the theorem is straightforward to generalize for the case when $f$ is only piecewise $C^1$. \qed

\smallskip

Two immediate corollaries of the Theorem \ref{tm:GRtm} in the $m=3$ case, applied to equation (\ref{eq:intPsi}), are original Runcorn's observations:

\begin{itemize}
\item[(1)] In the case of magnetization by \emph{internal} source, we take $\M(\vr') \defeq f(r') \nabla v(\vr')$ and $u(\vr') \defeq \norm{\vr - \vr'}^{-1}$, with fixed $\vr$ such that $r > b$; then $\Psi(\vr) = 0$ for $r > b$  and, consequently, magnetic field is zero in the exterior.

\item[(2)] In the case of magnetization by \emph{external} source, we take $\M(\vr') \defeq f(r') \nabla u(\vr')$ and $v(\vr') \defeq \norm{\vr - \vr'}^{-1}$, with fixed $\vr$ such that $r < a$; then $\Psi(\vr) = 0$ for $r < a$  and, consequently, magnetic field is zero in the interior.
\end{itemize}

One might ask whether a converse of the Theorem \ref{tm:GRtm} holds: If a function $f$ satisfies condition (\ref{eq:GR}) for all harmonic functions $u$ and $v$, satisfying conditions from the theorem, is $f$ necessarily isotropic, $f = f(\norm{\bm{x}})$? Discussion in Appendix B of \cite{AHD96} suggests that the answer is positive but, as far as we are aware of, there is no rigorous proof of this conjecture in the literature. Closely related question arises in the context of the inverse problems \cite{LV20}, reconstruction of the magnetization from the measured external field (cf.~\cite{PN10} for an analysis of the \emph{Lunar Prospector} data). This problem is more general in a sense that $\Omega$ does not have to be a spherical shell and magnetization does not have to be a multiple of gradient of harmonic function, but it is more special in a sense that one of the harmonic functions is fundamental solution for the Laplacian. One of the major obstacles here is the nonuniqueness which stems from the existence of ``magnetic annihilators'' \cite{MH03}, also referred to as ``invisible magnetizations'' \cite{BGKM21}, domains $\Omega$ with magnetizations $\M$ with magnetic field which vanishes in the whole or some of the connected components of the complement of $\Omega$.

%%%%%%%%%%%%%%%%%%%%%%%%%%%%%%%%%%%%%%%%%%%
\section{Approximate Runcorn's theorem} %%%
%%%%%%%%%%%%%%%%%%%%%%%%%%%%%%%%%%%%%%%%%%%

Now, we turn to various deviations from the initial setting of the Runcorn's theorem, when some of the geometric or analytic assumptions are relaxed. A direct step towards generalization could start with the pointwise Cauchy--Bunyakovsky--Schwarz inequality,
\be
|\nabla u(\bm{x}) \cdot \nabla v(\bm{x})| \le \norm{\nabla u(\bm{x})} \cdot \norm{\nabla v(\bm{x})}
\ee
and H\"older's inequality for three functions (cf.~chapter 2 in \cite{LiebLoss}), to obtain a bound
\be
\left| \int_{\Omega} f(\norm{\bm{x}}) \nabla u(\bm{x}) \cdot \nabla v(\bm{x}) \, \df V \right| \le \norm{f}_{L^{p}(\Omega)} \norm{\nabla u}_{L^q(\Omega)} \norm{\nabla v}_{L^r(\Omega)}
\ee
for all $p,q,r \in \left[1,\infty\right]$ such that $p^{-1} + q^{-1} + r^{-1} = 1$, assuming that all terms on the right-hand side are well-defined. However, from a physical point of view, it is more useful to look directly at the magnetic field, as an upper bound on the value of the scalar potential is not telling us much about its gradient. For simplicity (and clarity), we shall focus on the $m=3$ case only.

\smallskip

In the complement of the magnetized domain $\Omega$, for all $\vr \in \rr^3 - \overline{\Omega}$, field $\HH$ may be written directly from equation (\ref{eq:intPsi}), as
\be\label{eq:Bfar}
\HH(\vr) = \frac{1}{4\pi} \int_\Omega \M(\vr')\cdot\nabla' \frac{\vr - \vr'}{\norm{\vr-\vr'}^3} \, \df V' = \frac{1}{4\pi} \int_\Omega \frac{3(\bm{u}\cdot\M(\vr'))\bm{u} - \M(\vr')}{\norm{\vr-\vr'}^3} \, \df V' \, ,
\ee
with $\bm{u} \defeq (\vr - \vr')/\norm{\vr - \vr'}$. There are two basic problems that we may first treat separately: the one in which the function $\kappa$ is no longer isotropic and the one in which the crust is ``slightly deformed'' spherical shell.

%%%%%%%%%%%%%%%%%%%%%%%%%%%%%%%%%%%%%%%%%%%
\subsection{Anisotropic susceptibility} %%%
%%%%%%%%%%%%%%%%%%%%%%%%%%%%%%%%%%%%%%%%%%%

Let us assume that the crust is still a spherical shell $\mathcal{A} = \{ \bm{x} \in \rr^3 \mid a \le \norm{\bm{x}} \le b \}$, while the susceptibility $\kappa$ may be written as a sum of an isotropic function $\varsigma$ and its ``perturbation'' $\nu$,
\be
\kappa(\vr) = \varsigma(r) + \nu(\vr) \, ,
\ee
such that $|\nu(\vr)| \le \veps$ for all $\vr \in \mathcal{A}$ with some positive constant $\veps > 0$. As $\M = (\varsigma + \nu) \HH_0$, and we already know the null result via Theorem \ref{tm:GRtm} for $\varsigma$, it follows from (\ref{eq:Bfar}) that the magnetic field is
\be
\B(\vr) = \frac{\mu_0}{4\pi} \int_{\mathcal{A}} \nu(\vr') \, \frac{3(\bm{u}\cdot\HH_0(\vr'))\bm{u} - \HH_0(\vr')}{\norm{\vr-\vr'}^3} \, \df V' \, .
\ee
Schematically, we have an equation of the form
\be
\B(\vr) = \int_\mathcal{A} \bm{X}(\vr,\vr') \, \df V
\ee
and each component of the magnetic field may be bounded via
\be
|B^i(\vr)| = \left| \int_\mathcal{A} X^i(\vr,\vr') \, \df V' \right| \le \int_\mathcal{A} |X^i(\vr,\vr')| \, \df V' \le V(\mathcal{A}) \max_{\vr'\in\mathcal{A}} |X^i(\vr,\vr')| \, ,
\ee
where we have introduced the abbreviation $V(\mathcal{A}) = 4\pi(b^3 - a^3)/3$ for the volume of the spherical shell $\mathcal{A}$. Now, for any $\alpha > 2$, we have
\be
\norm{\alpha(\bm{u}\cdot\HH_0)\bm{u} - \HH_0}^2 = \alpha(\alpha-2)(\bm{u}\cdot\HH_0)^2 + \HH_0^2 \le (\alpha(\alpha-2) + 1)\HH_0^2 = (\alpha-1)^2 \HH_0^2 \, ,
\ee
with the equality holding at the given point iff $\HH_0$ and $\bm{u}$ are linearly dependent at that point. In our case, we have $\alpha = 3$, so that
\be
\norm{3(\bm{u}\cdot\HH_0(\vr'))\bm{u} - \HH_0(\vr')} \le 2\norm{\HH_0(\vr')} \le 2 H_0
\ee 
with $H_0 \defeq \max_{\vr'\in\mathcal{A}} \norm{\HH_0(\vr')}$. Finally, for any constant $0 < \delta < 1$ and any radius $r > b/\delta$ we have $(1 - b/r)^{-3} < (1 - \delta)^{-3}$. Hence, from the triangle inequality $\norm{\vr - \vr'} \ge r - r'$ and the assumption that $r' \le b < r$, it follows that 
\be
\frac{1}{{\norm{\vr-\vr'}^3}} \le \frac{1}{(r-b)^3} < \frac{1}{(1-\delta)^3} \, \frac{1}{r^3} \, .
\ee
Altogether, for any $r > b/\delta$, we have
\be\label{ineq:B1}
|B^i(\vr)| < \frac{C}{r^3} \qqd C \defeq \frac{\mu_0}{4\pi} \, 2H_0\veps \, \frac{V(\mathcal{A})}{(1-\delta)^3} \, .
\ee
As long as $\delta < 1/2$, corresponding to distances at least as twice the radius of the body, we have $(1 - \delta)^{-3} < 8$, so that this factor increases constant $C$ no more than one order of magnitude. However, as we approach the surface of the astrophysical body, the upper bound becomes less useful as the limit $\delta \to 1$ implies $C \to \infty$. An alternative approach is to use the formula, derived from (\ref{eq:altPsi}),
\be\label{eq:Bnear}
\HH(\vr) = \frac{1}{4\pi} \left( - \int_\Omega \frac{\nabla'\cdot\M(\vr')}{\norm{\vr-\vr'}^2} \, \bm{u} \, \df V' + \oint_{\dd\Omega} \frac{\M(\vr')\cdot\un}{\norm{\vr-\vr'}^2} \, \bm{u} \, \df a' \right) .
\ee
Even if we, for simplicity, assume that the magnetization is divergentless, $\nabla\cdot\M = 0$, remaining surface integral again lead us to an upper bound of the form $|B^i(\vr)| \le f(r)$, with some function $f$ for which the limit $\lim_{\veps \to 0^+} f(b+\veps)$ diverges. Intuitively, without any additional assumptions, one cannot exclude the possibility of magnetization ``concentrated'' around the point of the surface closest to the point where we observe the field.

%%%%%%%%%%%%%%%%%%%%%%%%%%%%%%%%%%%
\subsection{Nonspherical crust} %%%
%%%%%%%%%%%%%%%%%%%%%%%%%%%%%%%%%%%

\begin{figure}[ht!]
\centering
\includegraphics[]{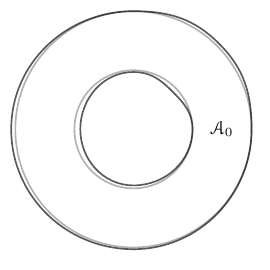}
\caption{Sketch of irregular crust (darker lines) and a maximal embedded spherical shell (lighter lines).} \label{fig:1}
\end{figure}

Let us now assume that the susceptibility $\kappa$ is isotropic, $\kappa = \kappa(r)$, while the crust is ``slightly deformed'' in a sense that the domain $\Omega$ is bounded by two smooth disjoint embedded surfaces $\mathcal{S}_-$ and $\mathcal{S}_+$, each homeomorphic to the 2-sphere, and admits embedding of a spherical shell $\mathcal{A}_0 = \{ \bm{x} \in \rr^3 \mid a \le \norm{\bm{x}} \le b \} \subseteq \Omega$, as sketched in Figure \ref{fig:1}. For the sake of better estimate, it is advantageous to choose spherical shell of volume as large as possible. We shall denote by $\Omega' \defeq \Omega - \mathcal{A}_0$ part of the domain which remains after we remove the spherical shell $\mathcal{A}_0$.

Again, as the Theorem \ref{tm:GRtm} gives the null result for the part of the integration over the spherical shell $\mathcal{A}_0$, we have immediately
\be
\B(\vr) = \frac{\mu_0}{4\pi} \int_{\Omega} \M(\vr')\cdot\nabla' \frac{\vr - \vr'}{\norm{\vr-\vr'}^3} \, \df V' = \frac{\mu_0}{4\pi} \int_{\Omega'} \frac{3(\bm{u}\cdot\M(\vr'))\bm{u} - \M(\vr')}{\norm{\vr-\vr'}^3} \, \df V' .
\ee
Using the same strategy as in the previous subsection, we can first introduce $R \defeq \max_{\vr\in\dd\Omega} \norm{\vr}$ and $M \defeq \max_{\vr\in\Omega'} \norm{\M(\vr)}$. The upper bound $M$ on magnetization may be estimated from the systematic study of samples from various sites on the astrophysical body. Then for any constant $0 < \delta < 1$ and any radius $r > R/\delta$, we have
\be\label{ineq:B2}
|B^i(\vr)| < \frac{C'}{r^3} \qqd C' \defeq \frac{\mu_0}{4\pi} \, 2M \, \frac{V(\Omega')}{(1-\delta)^3} \, ,
\ee
where $V(\Omega')$ is the volume of the domain $\Omega'$. An in-depth analysis of magnetic field for the particular case of ellipsoidal bodies was performed in \cite{JWL99}.

\smallskip

As a small addendum, we can provide a straightforward generalization of the results above, for the case when the susceptibility is anisotropic, $\kappa = \varsigma + \nu$, \emph{and} the crust is not spherical. Magnetic field has two contributions, one from the embedded spherical shell $\mathcal{A}_0$ with anisotropic part of susceptibility $\nu$ and one from aspherical part $\Omega'$ of the crust,
\be
\B(\vr) = \frac{\mu_0}{4\pi} \left( \int_{\mathcal{A}_0} \nu(\vr') \, \frac{3(\bm{u}\cdot\HH_0(\vr'))\bm{u} - \HH_0(\vr')}{\norm{\vr-\vr'}^3} \, \df V' + \int_{\Omega'} \frac{3(\bm{u}\cdot\M(\vr'))\bm{u} - \M(\vr')}{\norm{\vr-\vr'}^3} \, \df V' \right) .
\ee
Previous results directly lead to the bound
\be
|B^i(\vr)| < \frac{C+C'}{r^3}
\ee
for $r > R/\delta$, with all the notation (parameter $\delta$, radius $R$ and constants $C,C'$) defined as above.

%%%%%%%%%%%%%%%%%%%%%%%%%%%
\section{Layered crust} %%%
%%%%%%%%%%%%%%%%%%%%%%%%%%%

Even if we set aside corrections stemming from anisotropy and asphericity, Runcorn's model subtly relies on one more crucial assumption, ``instant'' acquisition of the magnetization throughout the whole crust. A more realistic scenario was analysed by Srnka \cite{Srnka76}, in which the thermoremanent magnetization is acquired as the astrophysical body cools inwards and magnetic field of each magnetized layer, in addition to the core's magnetic field, magnetizes newly cooled layers. Additional permeable effects were treated\footnote{This line of research builds on previous Stephenson's analysis \cite{Stephenson76a,Stephenson76b}. In the Stephenson's model, all layers, each with different permeability, are initially embedded in the core's magnetic field, and then they instantly acquire thermoremanent magnetization proportional to the ambient magnetic field, after which the core's magnetic field is ``turned-off''.} in the accompanied proceedings paper \cite{Srnka76p}, by taking into account permeability of the layers which were previously magnetized. 

\smallskip

Let us look more closely into the original Srnka's model. We go back to the simplest geometry, spherical body of radius $r_0$ and constant susceptibility $\kappa$, surrounded by vacuum. Although the atmosphere may, in principle, contribute to the magnetization (as an additional layer of permeable material), we shall neglect this effect\footnote{For example, lunar atmosphere is highly tenuous, with fewer than $10^6$ molecules per cubic centimetre. Investigations of the lunar atmosphere has a long and intriguing history, going back to Ru\dj er Bo\v skovi\'c's essay \emph{De lun\ae~atmosph\ae ra} \cite{RB1753}, over to modern surveys within numerous missions to the Moon \cite{Stern99}.}. Body's core contains a source of magnetic field $\B_\mathrm{c} = -\mu_0 \nabla \Psi_\mathrm{c}$, with magnetic scalar potential $\Psi_c$ which in the spherical coordinate system is given by
\be
\Psi_\mathrm{c}(r,\theta,\varphi) = \sum_{\ell,m} B_{\ell m}^{(\mathrm{c})} r^{-(\ell+1)} \fun{Y}_{\ell m}(\theta,\varphi) \, ,
\ee
with some fixed coefficients $B_{\ell m}^{(c)}$ (varying core's field could be implemented with coefficients $B_{\ell m}^{(c)}$ which change over the process of the crust's magnetization). In the first step, after the Curie's isotherm shrinks from $r = r_0$ to $r = r_1 < r_0$ sphere, spherical shell $r_1 < r < r_0$ acquires magnetization proportional to the core's magnetic field,
\be
\M_1(r,\theta,\varphi) = -\kappa \nabla \sum_{\ell,m} B_{\ell m}^{(\mathrm{c})} r^{-(\ell+1)} \fun{Y}_{\ell m}(\theta,\varphi) \, .
\ee
Magnetic field corresponding to this magnetized shell is, as a consequence of the generalized Runcorn's theorem, zero in the external region $r > r_0$, but nonzero in the interior region $r < r_1$ (details are derived in Appendix),
\be
\HH_1^{(\mathrm{in})}(r,\theta,\varphi) = \kappa \nabla \sum_{\ell,m} \frac{\ell+1}{2\ell+1} \, B_{\ell m}^{(\mathrm{c})} \big( r_1^{-(2\ell+1)} - r_0^{-(2\ell+1)} \big) r^\ell \fun{Y}_{\ell m}(\theta,\varphi)
\ee
Furthermore, after the Curie's isotherm shrinks from $r = r_1$ to $r = r_2 < r_1$ sphere, spherical shell $r_2 < r < r_1$ acquires magnetization proportional to the sum of the core's and the outermost shell's magnetic field, 
\be
\M_2(r,\theta,\varphi) = -\kappa \nabla \sum_{\ell,m} B_{\ell m}^{(\mathrm{c})} \left( r^{-(\ell+1)} - \kappa \, \frac{\ell+1}{2\ell+1} \big( r_1^{-(2\ell+1)} - r_0^{-(2\ell+1)} \big) r^\ell \right) \fun{Y}_{\ell m}(\theta,\varphi) \, .
\ee
It is important to note that magnetization $\M_2$ has two components: the one proportional to $r^{-(\ell+1)}$ (with corresponding magnetic field which is zero in $r > r_1$) and the one proportional to $r^\ell$ (with corresponding magnetic field which is zero in $r < r_2$). Thus, the next layer, spherical shell $r_3 < r < r_2$ will acquire magnetization proportional to the sum of the core's magnetic field and interior magnetic field of two outer layers,
\be
\M_3(r,\theta,\varphi) = -\kappa \nabla \sum_{\ell,m} B_{\ell m}^{(\mathrm{c})} \left( r^{-(\ell+1)} - \kappa \, \frac{\ell+1}{2\ell+1} \big( r_2^{-(2\ell+1)} - r_0^{-(2\ell+1)} \big) r^\ell \right) \fun{Y}_{\ell m}(\theta,\varphi) \, ,
\ee
and so on. If we split magnetization of each shell as follows, $\M_i = -\kappa\nabla\Psi_c + \M_i^{\mathrm{(eff)}}$, then the ``effective magnetization'' (part which contributes to the exterior magnetic field) of the $i$-th shell may be written as
\be
\M_i^{\mathrm{(eff)}} = \kappa^2 \nabla \sum_{\ell,m} B_{\ell m}^{(\mathrm{c})} \, \frac{\ell+1}{2\ell+1} \big( r_{i-1}^{-(2\ell+1)} - r_0^{-(2\ell+1)} \big) r^\ell \fun{Y}_{\ell m}(\theta,\varphi) \, ,
\ee
with the corresponding scalar potential in the external region
\be
\Psi_i^{\mathrm{(ext)}}(r,\theta,\varphi) = \kappa^2 \sum_{\ell,m} \frac{\ell(\ell+1)}{(2\ell+1)^2} \, B_{\ell m}^{(\mathrm{c})} \big( r_{i-1}^{-(2\ell+1)} - r_0^{-(2\ell+1)} \big) \big( r_{i-1}^{2\ell+1} - r_i^{2\ell+1} \big) r^{-(\ell+1)} \fun{Y}_{\ell m}(\theta,\varphi) \, .
\ee
Now, assuming that we have $N$ shells defined with the radii $0 < r_N < \dots < r_2 < r_1 < r_0$, total external magnetic field is given by superposition
\be
\B^{\mathrm{(ext)}} = -\mu_0 \nabla \sum_{i=1}^N \Psi_i^{\mathrm{(ext)}} \, .
\ee
Here we encounter the sum
\be
\sum_{i=1}^N \big( r_{i-1}^{-(2\ell+1)} - r_0^{-(2\ell+1)} \big) \big( r_{i-1}^{2\ell+1} - r_i^{2\ell+1} \big) = -1 + \left( \frac{r_N}{r_0} \right)^{\!2\ell+1} + \sum_{i=1}^N \left( 1 - \left( \frac{r_i}{r_{i-1}} \right)^{\!2\ell+1} \right) ,
\ee
which can be reorganized with a convenient abbreviation $x_i \defeq (r_{i-1} - r_i)/r_i$, with which $r_i/r_{i-1} = 1/(1 + x_i)$. Since we look at the finite number of shells, there are $x_m,x_M > 0$, such that $x_i \in [x_m,x_M]$ for all $i$. Finally, using the average layer thickness $d \defeq (r_0 - r_N)/N$ allows us to put strict bounds\footnote{Srnka \cite{Srnka76} introduces a number of approximations in the earlier stage of analysis and invokes inequality $r_0^2 d \ge (r_{i-1} - r_i) r_i^2 \ge r_N^2 d$, which does not hold in general, at least without any additional assumptions.},
\be
\frac{r_0 - r_N}{d} \left( 1 - \frac{1}{(1+x_m)^{2\ell+1}} \right) \le \sum_{i=1}^N \left( 1 - \left( \frac{r_i}{r_{i-1}} \right)^{\!2\ell+1} \right) \le \frac{r_0 - r_N}{d} \left( 1 - \frac{1}{(1+x_M)^{2\ell+1}} \right) .
\ee
Major obstacle here is the (lack of) knowledge about the average layer thickness $d$, as well as the constants $x_m$ and $x_M$ (in the lunar case Srnka argues for the estimate $d \sim 1\,$km). A simple shortcut is to use Taylor series $1 - (1 + \veps)^{-n} = n\veps + O(\veps^2)$ and rough estimates $x_m \sim d/r_0$ and $x_M \sim d/r_N$, in order to produce estimates for the lower bound $(2\ell+1)(1 - r_N/r_0)$ and the upper bound $(2\ell+1)(r_0/r_N - 1)$. These values, however, must be taken with a grain of salt, as any more credible bounds require further information about the interior structure of the crust.

%%%%%%%%%%%%%%%%%%%%%%%%%%%
\section{Final remarks} %%%
%%%%%%%%%%%%%%%%%%%%%%%%%%%

Runcorn's theorem is a hidden gem of magnetostatics, providing an important insight into the magnetic field of magnetized crusts of astrophysical bodies. In Section 2 we have offered one possible bird's-eye view on the mathematical origin of the Runcorn's result. Focal result here is Theorem 2.1, revealing specific orthogonality of gradients of harmonic functions on the spherical shell in the intersection of their domains. An immediate corollary is the vanishing of the exterior (interior) magnetic field for the spherical shell magnetized by the interior (external) source.

\smallskip

In Section 3 we have analysed several deviations of the Runcorn's theorem, coming from aniso\-tropic susceptibility and/or nonsphericity of the crust. Inequalities (\ref{ineq:B1}) and (\ref{ineq:B2}) quantify an intuitive expectation that the magnetic field in these cases is bounded by the dipole, $O(r^{-3})$ term. Let us look briefly at prospect of more concrete estimates implied by these results. Global orbital measurements of magnetic field for some astrophysical bodies, such as the Moon (\emph{Lunar Prospector} mission \cite{LP08}) and Mars (\emph{Mars Global Surveyor} and \emph{MAVEN} missions \cite{MJ22}), are already at our disposal. Volume $V(\mathcal{A})$ can be inferred from seismographic data: e.g.~average thickness of the lunar crust \cite{Wieczorek09} is $\sim 40\,$km, while average thickness of the Martian crust \cite{Kim23} is $\sim 50\,$km. However, crucial obstacle for the evaluation of the constant $C$ in (\ref{ineq:B1}) arises with the difficulty of determination of the magnetizing field $H_0$. Setting aside rough paleofield estimates from meteorites of known origin, even when we have rock samples (at the time of writing this paper, no samples have been brought back from Mars), their analysis is burdened with methodological uncertainties \cite{WT14}. We can bring conclusions about the constant $C'$ in (\ref{ineq:B2}) one step further, at least for the Moon. Lunar surface may, at the first order of approximation, be described as an oblate ellipsoid with the polar radius $R_p \approx 1736\,$km and equatorial radius $R_e \approx 1738\,$km. Given that we assume that the inner surface of lunar crust, again at the lowest approximation, is a sphere, we have $V(\Omega')_\mathrm{Moon} \approx 4\pi R_p(R_e^2 - R_p^2)/3$. Furthermore, using the estimate $M \sim 10^{-2}\,$Am$^{-1}$ \cite{CWPH11}, it follows that $(1-\delta)^3 C'_\mathrm{Moon} \sim 0.05\,$nT$\,(10^6\,$m$)^3$ (we are using cubic megametres as the radial distance will be usually expressed in thousands of kilometres).

\smallskip

Finally, in Section 4 we have revisited a model of a crust magnetized during the inward cooling, with the sharpening of the Srnka's estimates. It would be interesting to find a generalization of the latter model, to a crust which is magnetized during the continuous process of inward cooling. Also, one could take into account various external magnetic fields, mentioned in the discussion of the generalized Runcorn's Theorem 2.1, and study their effect on the crustal magnetization.

%%%%%%%%%%%%%%%%%%%%%%%%%%%%%%%%%%%%%%%%%%
\section*{Data availability statement} %%%
%%%%%%%%%%%%%%%%%%%%%%%%%%%%%%%%%%%%%%%%%%

No new data were created or analysed in this study. Data sharing is not applicable to this article.

%%%%%%%%%%%%%%%%%%%%%%%%%%%%%%%%%%%%%%%%%%%%%%%%%%%%%%%%%%
%%%%%%%%%%%%%%%%%%%%%%%%%%%%%%%%%%%%%%%%%%%%%%%%%%%%%%%%%%
\section*{Acknowledgements}

The research was supported by the Croatian Science Foundation Project No.~IP-2020-02-9614.

%%%%%%%%%%%%%%%%%%%%%%%%%%%%%%%%%%%%%%%%%%%%%%%%%%%%%%%%%%
%%%%%%%%%%%%%%%%%%%%%%%%%%%%%%%%%%%%%%%%%%%%%%%%%%%%%%%%%%

%%%%%%%%%%%%%
%%%%%%%%%%%%%
\appendix %%%
%%%%%%%%%%%%%
%%%%%%%%%%%%%

%%%%%%%%%%%%%%%%%%%%%%%%%%%%%%%%%%%
\section{Two auxiliary results} %%%
%%%%%%%%%%%%%%%%%%%%%%%%%%%%%%%%%%%

An elementary problem that was used in Section 4 consists of a single spherical shell $r_- < r < r_+$ with magnetization $\M$, surrounded by vacuum. Magnetic scalar potential, introduced via $\HH = -\nabla \Psi$, is a harmonic function, which in the spherical coordinate system may be written in the form
\be
\Psi(r,\theta,\varphi) = \left\{ \begin{array}{ll} \sum_{\ell,m} B_{\ell m}^{\mathrm{(ext)}} r^{-(\ell+1)} \fun{Y}_{\ell m}(\theta,\varphi) \, , & r_+ \le r \\[0.75em] \sum_{\ell,m} (A_{\ell m} r^\ell + B_{\ell m} r^{-(\ell+1)}) \fun{Y}_{\ell m}(\theta,\varphi) \, , & r_- \le r \le r_+ \\[0.75em] \sum_{\ell,m} A_{\ell m}^{\mathrm{(in)}} r^\ell \fun{Y}_{\ell m}(\theta,\varphi) \, , & r \le r_- \end{array} \right.
\ee
Coefficients $B_{\ell m}^{\mathrm{(ext)}}$, $A_{\ell m}$, $B_{\ell m}$ and $A_{\ell m}^{\mathrm{(in)}}$ may be found from the junction conditions, continuity of the potential $\Psi$ (in the absence of free currents) and continuity of the normal component of the magnetic field $\B = \mu_0(\HH + \M)$ on spheres $r = r_-$ and $r = r_+$. We shall consider two separate subcases:

\begin{itemize}
\item[(a)] Magnetization is of the form $\M = -\nabla \sum_{\ell,m} M_{\ell m} r^{-(\ell+1)} \fun{Y}_{\ell m}$. Solution of junction conditions is
\be
B_{\ell m}^{\mathrm{(ext)}} = 0 \qqd A_{\ell m} = \frac{\ell+1}{2\ell+1} \, M_{\ell m} r_+^{-(2\ell+1)} \qqd B_{\ell m} = -\frac{\ell+1}{2\ell+1} \, M_{\ell m} \, ,
\ee
\be
A_{\ell m}^{\mathrm{(in)}} = -\frac{\ell+1}{2\ell+1} \, M_{\ell m} (r_-^{-(2\ell+1)} - r_+^{-(2\ell+1)}) \, .
\ee
Vanishing of the exterior magnetic potential, $B_{\ell m}^{\mathrm{(ext)}} = 0$, is a direct consequence of the generalized Runcorn's theorem.

\smallskip

\item[(b)] Magnetization is of the form $\M = -\nabla \sum_{\ell,m} M_{\ell m} r^\ell \fun{Y}_{\ell m}$. Solution of junction conditions is
\be
A_{\ell m}^{\mathrm{(in)}} = 0 \qqd A_{\ell m} = -\frac{\ell}{2\ell+1}\,M_{\ell m} \qqd B_{\ell m} = \frac{\ell}{2\ell+1}\,M_{\ell m} r_-^{2\ell+1} \, ,
\ee
\be
B_{\ell m}^{\mathrm{(ext)}} = -\frac{\ell}{2\ell+1} \, M_{\ell m} (r_+^{2\ell+1} - r_-^{2\ell+1}) \, .
\ee
Vanishing of the interior magnetic potential, $A_{\ell m}^{\mathrm{(int)}} = 0$, is a direct consequence of the generalized Runcorn's theorem.
\end{itemize}

%%%%%%%%%%%%%%%%%%%%%%%%%%%%%%%%%%%
\bibliographystyle{amsalpha}%%%%%%%
%%%%%%%%%%%%%%%%%%%%%%%%%%%%%%%%%%%
\bibliography{m}%%%%%%%%%%%%%%%%%%%
%%%%%%%%%%%%%%%%%%%%%%%%%%%%%%%%%%%

\end{document}